\documentclass[10pt, conference, letterpaper]{IEEEtran}
\usepackage{float}
\usepackage{amsmath}
\usepackage{amssymb}
\usepackage{amsthm}
\usepackage{multirow}
\usepackage{tabularx}
\usepackage{graphicx}
\usepackage{subfigure}
\usepackage{authblk}
\usepackage[linesnumbered, ruled]{algorithm2e}

\usepackage{threeparttable}
\usepackage{slashbox}
\usepackage{algorithmic}
\usepackage{tabularx}
\usepackage{booktabs}
\usepackage{amsfonts}
\usepackage{dsfont}
\usepackage[all]{xy}

\def\squarebox#1{\hbox to #1{\hfill\vbox to #1{\vfill}}}

\newtheorem{theorem}{Theorem}

\newtheorem{definition}{Definition}

\begin{document}

\title{CIUV: Collaborating Information Against\\ Unreliable Views}

\author{Zimu Yuan$^{\dagger\ddagger}$, and Zhiwei Xu$^{\ddagger}$  \\ 
$^{\dagger}$University of Chinese Academy of Sciences, China\\
$^{\ddagger}$Institute of Computing Technology, Chinese Academy of Sciences, China\\
\{yuanzimu, zxu\}@ict.ac.cn
}

\maketitle \thispagestyle{empty}

\begin{abstract}
In many real world applications, the information of an object can be obtained from multiple sources. The sources may provide different point of views based on their own origin. As a consequence, conflicting pieces of information are inevitable, which gives rise to a crucial problem: how to find the truth from these conflicts. Many truth-finding methods have been proposed to resolve conflicts based on information trustworthy (i.e. more appearance means more trustworthy) as well as source reliability. However, the factor of men's involvement, i.e., information may be falsified by men with malicious intension, is more or less ignored in existing methods. Collaborating the possible relationship between information's origins and men's participation are still not studied in research. To deal with this challenge, we propose a method -- Collaborating Information against Unreliable Views (CIUV) --- in dealing with men's involvement for finding the truth. CIUV contains 3 stages for interactively mitigating the impact of unreliable views, and calculate the truth by weighting possible biases between sources. We theoretically analyze the error bound of CIUV, and conduct intensive experiments on real dataset for evaluation. The experimental results show that CIUV is feasible and has the smallest error compared with other methods.
\end{abstract}

\section{Introduction}
One significant challenge of Big Data is its wide variety, i.e., one can find the information about an object from various sources. The widely distributed sources may provide different point of views based on their origin or bias. When one want to find the truth out of these views, he should be able to determine how these conflicts come from, and which answers are more reliable than others. To deal with the challenge of variety, techniques have been proposed to assess the reliability of sources and derive the truth out of the information provided by sources \cite{1_1}\cite{1_2}\cite{1_3}\cite{1_4}\cite{1_5}. A common way of these techniques is to average among sources where they vote each other to assign weights or are weighted by a third-party.

Although many truth-finding techniques have been proposed, the factor of men's involvement more or less slip researchers' mind. For instance, someone may provide bias views with malicious intention so as to prevent others finding the truth; men are not incentivized to improve the reliability of the data they provide. Collaborating these possibilities are still not studied in research. Meanwhile, it has great application value to look up into the factor of men's involvement for truth-finding from multiple sources. A typical such application scenario is the Gross Demestic Product (GDP) accounting. There is a constant and ongoing debate about the accuracy of the GDP statistics. Men argue that GDP are overestimated or underestimated based on their bias observation, such as companies deliberately record less transaction amount on their books for paying less tax, or exaggerate the transaction amount for getting more loans. With the doubt in mind, some people start to assess the accuracy of the GDP statistics through data obtained from other more reliable sources, e.g. using the railway cargo volume, electricity consumption and loan disbursed by banks as indicators of GDP \cite{KeqiangIndex}. Those data provided by reliable sources are thought to be less manipulated by men, and people usually have more confidence on their accuracy. As can be seen, the key point of GDP accuracy lies in the reliability of sources. In many cases, men's malicious intention or negligence would make a source unreliable.

In this paper, we address the challenge of finding the truth with the consideration of men's involvement. In particular, we are interested in dealing with unreliable views that are provided by men with malicious intention. In summary, we make the following contributions in this paper:
\begin{itemize}
  \item We model the problem of finding the truth with men's involvement, and propose the method of Collaborating Information against Unreliable Views (CIUV), divided in 3 stages for accuracy improvement.
  \item CIUV interactively mitigates the impact of unreliable views, and calculates the truth by considering both mean and variance of error of views. We theoretically analyze the error bound of CIUV.
  \item In experiment, we evaluate the performance of CIUV with real dataset under various settings. By comparison, the experimental results show that CIUV is feasible and has the smallest error.
\end{itemize}

The remainder of this paper is organized as follows. We introduce the preliminary notations and their definitions in Section \ref{preliminaries}. We give an overview on the workflow of CIUV in Section \ref{overview}. We detail the 3 stages of CIUV implementation in Section \ref{CIUV_implementation}. We analyze the error bound of CIUV in Section \ref{error_analysis}. We provide the experimental results in Section \ref{experiment}. We review related work in Section \ref{related_work}. Finally, we conclude our paper in Section \ref{conclusion}.

\section{Preliminaries} \label{preliminaries}
We first introduce notations and their definitions, and then specify the objective in this section.

Suppose that there are total $m$ men providing the information about an object (i.e., the statistics data on GDP) separately. Let $V=\{v_i|i=1,2,...,m\}$ denote the set of views of these information. These views may be in different representations. For instance, someone may give the data of railway cargo volume while another person contribute the data of electricity consumption in last several years. As known, both of these two types of data can be used as indicators of GDP.
We map the view of different representations into an unified one. Let $U$ be the possible view space of the unified representation, and $M$ be the collection of mapping functions. For any $v\in V$, there exists a corresponding $u\in U$, and we can construct a mapping function $m\in M$ for $v$ to $u$. The formal definition of mapping function is given as follow.
\begin{definition}
A mapping function $m\in M: \{v\} \rightarrow \{u\}$ converts the view $v \in V$ provided by a man to the view $u \in U$ of unified representation.
\end{definition}

To evaluate the difference between any $u_i \in U$ and $u_j \in U$, we define a distance function $d(u_i,u_j)$ between them. By specifying the distance, we can distinguish how differences two views are, and further pick out the unreliable views. The formal definition of distance function is given as follow.
\begin{definition}
The distance function $d: U \times U \rightarrow R^+$ uses a positive real number to represent the difference between any $u_i,u_j \in U$.
\end{definition}
For simplicity, we use $u_i-u_j$ to represent $d(u_i,u_j)$ hereafter. In addition, also note that $u_i+u_j$ and $|u_i|$ are equal to $d(u_i,-u_j)$ and $d(u_i,0)$ respectively.

Given $U$, our object is to determine which views are reliable or unreliable, and estimate a truth view $u^*$ as the output result. In detail, we are interested in quantifying the reliability of views, and based on it, averaging among weighted views. Let $u^g$ be the ground truth. The formal definition of $u^*$ is given as follow.
\begin{definition}
The truth view
\begin{equation} \label{e_truth}
u^{*}=\frac{\sum_{i=1}^{m}w_i u_i}{\sum_{i=1}^{m}w_i}
\end{equation}
where the weights $w_i$, $i=1,2,...,m$ are assigned to views respectively. By weight assignment, $u^*$ is supposed to be the best estimated view approaching the ground truth $u^g$.
\end{definition}

\section{Overview} \label{overview}

\begin{figure}
\centering
\includegraphics[width=3.4in]{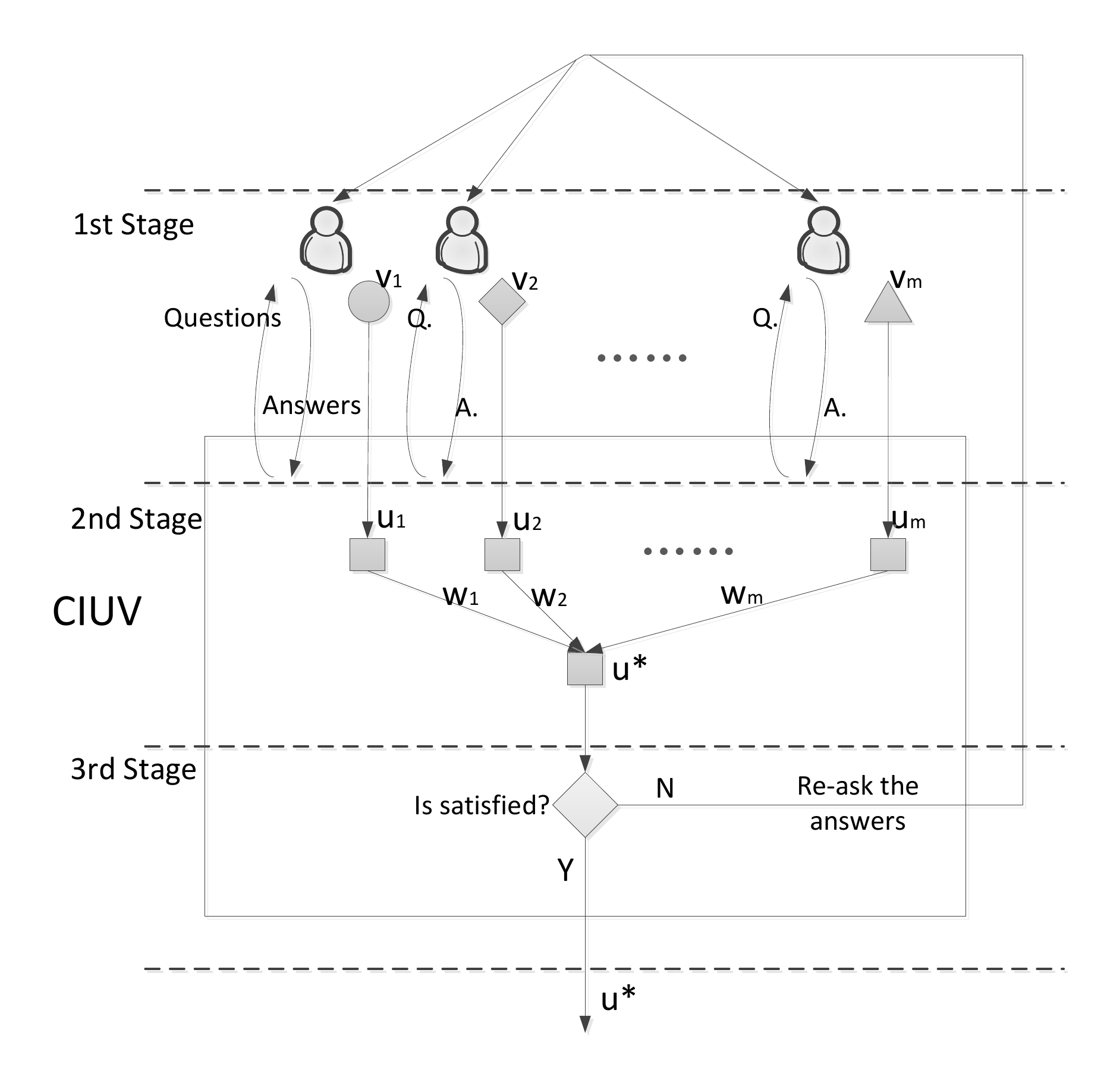}\\
\caption{\textrm{The workflow of CIUV.}} \label{fig_workflow}
\end{figure}

The workflow of CIUV is shown in Figure \ref{fig_workflow}. It includes 3 stages.
\begin{itemize}
  \item 1st stage: CIUV is interested in finding the truth $u^*$, but he does not intend to expose his purpose. So CIUV first asks a series of questions besides the questions about $u^*$ as masks. Then the men provide their views as answers for these series of questions. Next CIUV determines the reliability of the views based on the ground truths he already knows or by checking the consistency of the answers a man provides.
  \item 2nd stage: CIUV quantifies the reliability of views, and assigns greater weight to more reliable views and lighter weight to unreliable views. Then, CIUV averages on these weighted views to calculate a truth view $u^*$ as the output result.
  \item 3rd stage: If CIUV is not satisfied with the result $u^*$, he may re-ask the answers from the men by means of incentivization or punishment. CIUV would repeat this process until he get a acceptable result or the accuracy of the result is hardly improved.
\end{itemize}

\section{CIUV Implementation} \label{CIUV_implementation}
We introduce CIUV following the sequence of its 3 stages. In the 1st stage, CIUV determines the reliability of views provided by men; in the 2nd stage, CIUV calculate a truth $u^*$ by weighting views; in the 3rd stage, CIUV improves the accuracy of calculated $u^*$ by means of incentivization or punishment.

\subsection{1st stage}
In the 1st stage, CIUV aims to determine the reliability of views by asking a series of questions. There are two types of question should be considered. The first type is its ground truth is already known. For this type, CIUV first assesses the reliability by deriving the mean error $\mu$ and variance $\sigma^2$ of views provided by men. Let $S$ denote the set of views corresponding to the series of questions. For the $i$th man, we have
\begin{equation}
\mu_i = \frac{\sum_{s\in S} (s^g-s_i)}{|S|}
\end{equation}
\begin{equation}
\sigma_i^2 = \frac{\sum_{s\in S} (\mu_i-s_i)^2}{|S|}
\end{equation}
where, as defined in Section \ref{preliminaries}, $s^g-s_i$ (a simple representation of $d(s^g,s_i)$) is the distance between the ground truth $s^g$ and the view $s_i$ provided by the $i$th man. The second type of questions (including $u^*$) is its ground truth $s^g$ is still unknown. In this case, CIUV could use mean value of $s_i$, $i=1,2,...,m$ \cite{ra_1} or average over historical reliability values \cite{1_5} as the truth $s^*$, and then infer $\mu$ and $\sigma^2$.

CIUV assumes that men provides their views independently, i.e., they does not collude to provide a same bias view. Otherwise, we can divide men into independent groups such that any two men belong different groups give unrelated views for CIUV. Since it is not the major concern in our paper, its discussion is omitted here. Thereupon, we can use Gaussian distribution to describe the error (or to say reliability) of views in CIUV. For the $i$th man, we have
\begin{equation}
e_i \sim G(\mu_i, \sigma_i^2)
\end{equation}

\subsection{2nd stage}
Here we describe how to derive the truth $u^*$ in CIUV. The basic idea of our approach is that the reliable views are provided by reliable men, the ones with small $\mu$ and simultaneously with small $\sigma$. In other words, the truth $u^*$ should be close to views that are more trustworthy. We apply the weighted averaging strategy (Equation (\ref{e_truth})) to calculate $u^*$. The critical point lies in it is how to assign weights $w_1,w_2,...,w_m$ to views $u_1,u_2,...,u_m$.

With the assumption that men provides their views independently, we have the error $e^*$ of the truth $u^*$ following Gaussian distribution:
\begin{equation}
e^* \sim G(\frac{\sum_{i=1}^{m}w_i \mu_i}{\sum_{i=1}^{m}w_i}, \frac{\sum_{i=1}^{m}w^2_i \sigma^2_i}{(\sum_{i=1}^{m}w_i)^2})
\end{equation}
Without loss of generality, we restrict $\sum_{i=1}^{m}w_i=1$. Suppose that we have a error threshold value $e_T$. The objective of CIUV is to maximize the probability $P(|e^*|<e_T)$:
\renewcommand{\arraystretch}{1.5}
\begin{equation} \label{e_optimal}
    \begin{array}{ll}
        \max \; {P(|e^*|<e_T)}\\
        s.t. \; \sum_{i=1}^{m}w_i=1, w_i \geq 0.
    \end{array}
\end{equation}
\renewcommand{\arraystretch}{0.667}

\begin{figure}
\centering
\includegraphics[width=1.8in]{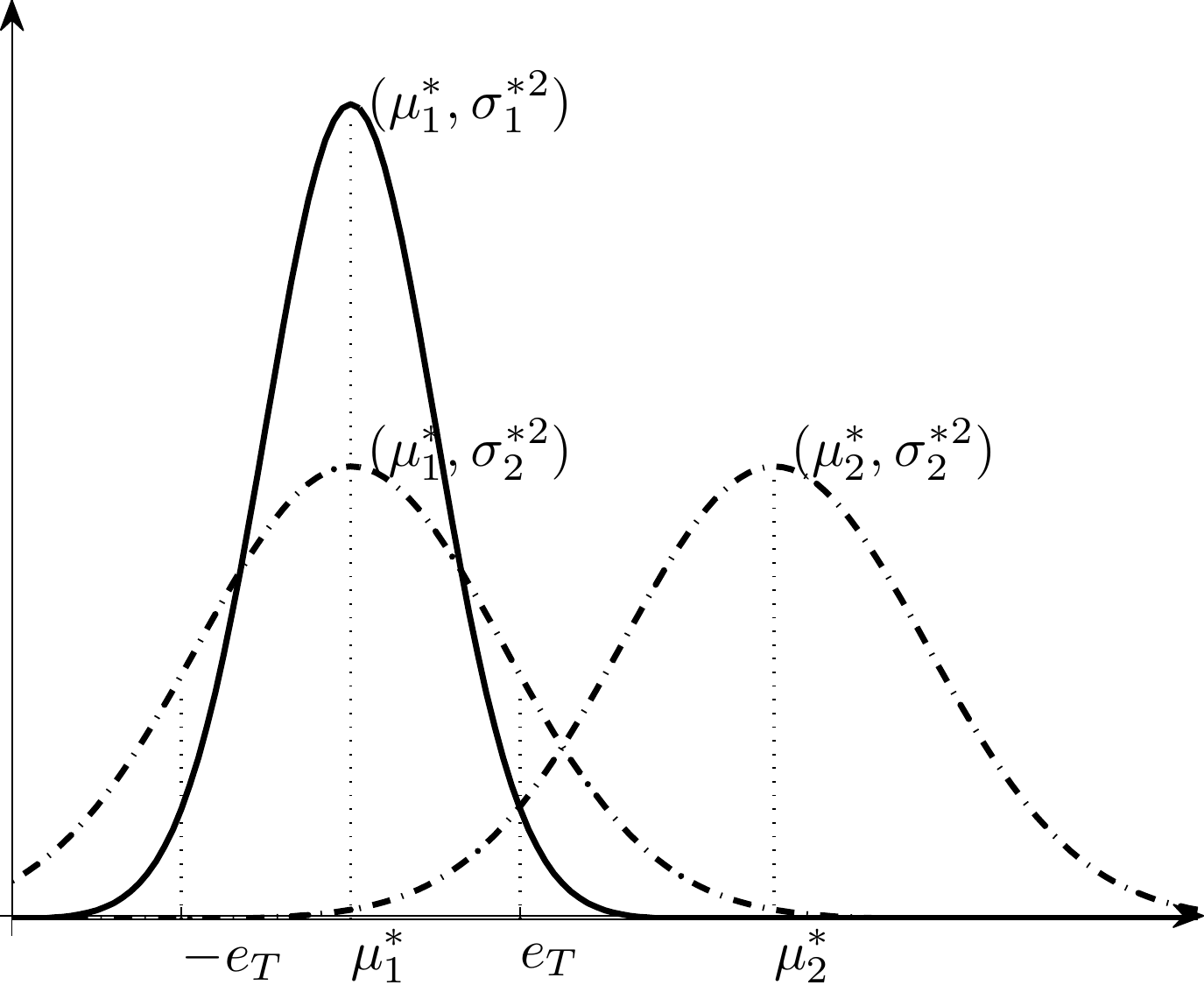}\\
\caption{\textrm{An example on Gaussian distribution.}} \label{fig_gauss}
\end{figure}

Let $\mu^*=\frac{\sum_{i=1}^{m}w_i \mu_i}{\sum_{i=1}^{m}w_i}$ and $\sigma^*=\frac{\sum_{i=1}^{m}w^2_i \sigma^2_i}{(\sum_{i=1}^{m}w_i)^2}$. In Figure \ref{fig_gauss}, we have $\mu^*_1<\mu^*_2$ and $\sigma^*_1<\sigma^*_2$. As can be seen, under the Gaussian distribution $G(\mu^*_1, \sigma^{*2}_1)$, $P(|e^*|<e_T)$ has a greater value compared with the other two distributions. In other words, to maximize $P(|e^*|<e_T)$, CIUV should weight views with the smallest combination of $|\mu^*|$ and $\sigma^*$. Substituting Gaussian probability-density function in $P(|e^*|<e_T)$, we have
\begin{equation} \label{e_prob}
P(|e^*|<e_T)=\int_{-e_T}^{e_T} \frac{1}{\sqrt{2\pi}\sigma^*} Exp(-\frac{(e^*-\mu^*)}{2\sigma^{*2}}) \; d\,e^*
\end{equation}
Unfortunately, it can be verified that the objective of maximizing $P(|e^*|<e_T)$ with Equation (\ref{e_prob}) cannot be directly solved. As a consequence, we are unable to derive the smallest combination of $\mu^*$ and $\sigma^*$ straightly.

Instead, we apply an approximate measure by firstly dealing with $\mu^*$ and $\sigma^*$ separately. We assume $e^*_{\mu} \sim G(\mu^*, 0)$ and $e^*_{\sigma} \sim G(0, \sigma^{*2})$. Then, the objective of CIUV (Equation (\ref{e_optimal})) is turned into $\max \; {P(|e^*_{\mu}|<e_T)}$ and $\max \; {P(|e^*_{\sigma}|<e_T)}$ respectively. Both of these two optimization problems are convex, so we can find global minimum guarantees of $e^*_{\mu}$ and $e^*_{\sigma}$ with the best weight assignment \cite{weight}. For $e^*_{\mu}$, it has many feasible solutions of maximizing $P(|e^*_{\mu}|<e_T)$. So we adopt a simple solution of weight assignment with the intuition that a $\mu_i \rightarrow 0$ should be assigned a greater weight, otherwise assigned a lighter weight:
\begin{equation}
w_{\mu,i} \varpropto \frac{1}{|\mu_i|}
\end{equation}
For $e^*_{\sigma}$, its closed form solution is:
\begin{equation}
w_{\sigma,i} \varpropto \frac{1}{\sigma^2_i}
\end{equation}
We normalize both weight assignments with the constraint of $\sum_{i=1}^{m} w_{\mu,i}=1$ and $\sum_{i=1}^{m} w_{\sigma,i}=1$ respectively. For $e^*_{\mu}$, its weight assignment is:
\begin{equation}
w_{\mu,i}=\frac{\prod_{k=1,k\neq i}^m u_k}{\sum_{j=1}^{m} \prod_{k=1,k\neq j}^m u_k}
\end{equation}
For $e^*_{\sigma}$, its weight assignment is:
\begin{equation}
w_{\sigma,i}=\frac{\prod_{k=1,k\neq i}^m \sigma^2_k}{\sum_{j=1}^{m} \prod_{k=1,k\neq j}^m \sigma^2_k}
\end{equation}
Based on the above weight assignments of $\max{\; {P(|e^*_{\mu}|<e_T)}}$ and $\max \; {P(|e^*_{\sigma}|<e_T)}$, we designate the weights for the problem of maximizing $P(|e^*|<e_T)$ by the following proportion:
\begin{equation}
w_i \varpropto \min\{w_{\mu,i}, w_{\sigma,i}\}
\end{equation}
Normalizing the weights with the constraint of $\sum_{i=1}^{m}w_i=1$, we have
\begin{equation}\label{e_weight}
w_i = \frac{\min\{w_{\mu,i}, w_{\sigma,i}\}}{\sum_{i=1}^m \min\{w_{\mu,i}, w_{\sigma,i}\}}
\end{equation}

CIUV adopts the weight assignment of Equation (\ref{e_weight}). Its advantage is that a view with either great $\mu^*$ or great $\sigma^*$ would be designated a light weight, and then we will have more confidence on the accuracy of the final output view $u^*$.

\subsection{3rd stage}
CIUV prompts the accuracy of the output view $u^*$ by means of incentivization or punishment. In implementation, CIUV does not care about the details about the adopted means. Instead, it assumes that the adopted means is effective, and focuses on the dispatch of accuracy improvement. Here, the effective means imply that each iteration of CIUV process will give more confidence on the accuracy of the output view $u^*$. The formal definition is as follow:
\begin{definition}[\textmd{Effective Means}]
With the means adopted, the $(j+1)$th iteration of CIUV process has the maximum probability $P(|e^*(j+1)|<e_T)$ compared with previous iterations, such as $P(|e^*(j+1)|<e_T) \geq P(|e^*(j)|<e_T)$
\end{definition}
In other words, the adopted means is effective in that they stimulate the men motivated to provide more trustworthy views.

At the beginning of process, CIUV sets an infinite error value to the views $u_1(0),u_2(0),...,u_m(0)$, namely $e_1(0)=e_2(0)=...=e_m(0)=\infty$; sets two threshold values: acceptable confidence probability $R$, i.e., $P(|e^*(j+1)|<e_T)\geq R$, on error, and lower bound of confidence improvement $D$, i.e., $P(|e^*(j+1)|<e_T)-P(|e^*(j)|<e_T)\geq D$. Then, at each iteration, CIUV verifies if $R$ is reached or can hardly be reached (namely the confidence improvement is lower than $D$ between two iterations in front and back).

\begin{algorithm} \label{flow_of_CIUV}
\scriptsize
\caption{CIUV Process}
\KwOut{The output truth view $u^*$}
Set $j=-1$, $e^*(0)=\infty$, $e_i(0)=\infty$, $i=1,2,...,m$;\\
\Repeat{1}{
    Set $j=j+1$;\\
    Test the reliability of views, and use Gaussian distribution $G(\mu_i,\sigma_i^2)$, $i=1,2,...,m$, to describe the error of views;\\
    Assign weight $w_i = \frac{\min\{w_{\mu,i}, w_{\sigma,i}\}}{\sum_{i=1}^m \min\{w_{\mu,i}, w_{\sigma,i}\}}$ (Equation (\ref{e_weight})), $i=1,2,...,m$, to views;\\
    Calculate the view $u^*(j+1)=\frac{\sum_{i=1}^{m}w_i u_i}{\sum_{i=1}^{m}w_i}$ (Equation (\ref{e_truth}));\\
    \eIf{$P(|e^*(j+1)|<e_T)\geq R$ or $P(|e^*(j+1)|<e_T)-P(|e^*(j)|<e_T)< D$}{
        Set $u^*=u^*(j+1)$;\\
        \Return $u^*$;\\
    }{
        Simulate the $i$th man to provide more trustworthy view if $P(|e_i(j+1)|<e_T)-P(|e_i(j)|<e_T)\geq D$, $i=1,2,...,m$;\\
    }
}
\end{algorithm}

Assume that CIUV process is in $(j+1)$th iteration. CIUV firstly checks whether $P(|e^*(j+1)|<e_T)\geq R$ or $P(|e^*(j+1)|<e_T)-P(|e^*(j)|<e_T)< D$ is satisfied. If so, it outputs the final view $u^*(j+1)$ by the weight assignment of $w_1,w_2,...,w_m$ in Equation (\ref{e_weight}). Else, it checks the confidence improvement of each view provided by men separately. In detail, for the $i$th man, CIUV calculates the probability $P(|e_i(j+1)|<e_T)$, subject to $w_i=1$ and $w_k=0, k\neq i, k=1,2,...,m$. If $P(|e_i(j+1)|<e_T)-P(|e_i(j)|<e_T)<D$, CIUV believes that it is hard to stimulate the $i$th man to provide more trustworthy views, and stops the effect on him; else CIUV continues to stimulate this man by means of incentivization or punishment in the next iteration.

\subsection{summary}
Here we summarize the overall flow of CIUV process in Algorithm \ref{flow_of_CIUV}.
The time complexity of the CIUV process is in linear with the total number of $m$ views provided by men, i.e. $O(Cm)$, where $C$ is related to the number of iterations, the time complexity on testing the reliability of views, and the time complexity on calculating the truth view $u^*$.

\section{Error Analysis} \label{error_analysis}
Consider the worst case of the error $e^*$ on $u^*$. Let the ground truth view $u^g=\sum_{i=1}^m w_i^g u_i$ (Equation (\ref{e_truth})), subject to $\sum_{i=1}^m w^g_i=1$, while $u^*=\sum_{i=1}^m w_i u_i$, subject to $\sum_{i=1}^m w_i=1$. Without loss of generality, assume that the worst case is the distance $e^*=u^g-u^*$ reaches the maximal value. We have
\begin{equation}\label{e_inequal}
u^g-u^* \leq \sum_{i=1}^m (w^g_i-w_i)|u_i|
\end{equation}
where $|u_i|$ is equal to $d(u_i,0)$ as stated in Section \ref{preliminaries}. Let $|u_{max}|=\max \{|u_1|,|u_2|, ...,|u_m|\}$. We have the following theorem:
\begin{theorem}
The view $u^*$ output by CIUV is in distance of $|u_{max}|$ to the ground truth $u^g$ in worst case.
\end{theorem}
\begin{proof}
For the $i$th view, we have either $w^g_i\geq w_i$ or $w^g_i< w_i$. Taking out the negative parts, $w^g_i-w_i<0$, $i=1,2,...,m$, we can rewrite Inequality (\ref{e_inequal}) as:
\renewcommand{\arraystretch}{1.5}
\begin{equation}
\begin{array}{ll}
u^g-u^* & \leq \sum_{i=1,w^g_i\geq w_i}^m (w^g_i-w_i)|u_i| \\
& \leq |u_{max}|\sum_{i=1,w^g_i\geq w_i}^m (w^g_i-w_i) \\
\end{array}
\end{equation}
\renewcommand{\arraystretch}{0.667}
Considering $\sum_{i=1,w^g_i\geq w_i}^m w^g_i=1$ and $\sum_{i=1,w^g_i\geq w_i}^m w_i=1$, we have $u^g-u^* \leq |u_{max}|$ at worse.
\end{proof}

Besides, we can easily see $u^*=u^g$ without error at best.

\section{Experiment} \label{experiment}

In this section, we give experimental results on GDP dataset under various scenarios. The experimental results show the higher accuracy of CIUV compared with other methods. Next we first introduce the experimental setting in Section \ref{experiment_setting}, and then conduct the experiments on varying malicious views, varying manipulation factors, varying improvement factor in Section \ref{experiment_mv}, \ref{experiment_mf}, and \ref{experiment_if} respectively.

\begin{table} \scriptsize
\renewcommand{\arraystretch}{1.0}
\caption{Views on GDP growth rate} \label{table_GDP_growth_rate}
\centering
\begin{tabular}{p{1.2cm}p{3.5cm}p{1cm}p{1.2cm}}
  \hline
  Views & Full Name & Mean Error & Standard Deviation \\
  \hline
  FCE & Final Consumption Expenditure & 2.4069 & 1.5291 \\
  GCF & Gross Capital Formation & 3.8193 & 2.9389 \\
  NE & Net Exports & 33.6287 & 34.5794 \\
  GDP\_EA & GDP by The Expenditure Approach & 1.2462 & 0.9685 \\
  NPT & Net Production Tax & 3.9390 & 3.4461 \\
  WC & Worker Compensation & 4.1153 & 5.3371 \\
  DFA & Depreciation of Fixed Assets & 3.6984 & 2.3672 \\
  BB & Business Balance & 10.7253 & 14.3010 \\
  GDP\_IA & GDP by The Income Approach & 3.0893 & 3.6595 \\
  FI & GDP of The First Industry & 4.8382 & 3.2961 \\
  SI & GDP of The Secondary Industry & 1.6570 & 1.1663 \\
  TI & GDP of The Tertiary Industry & 2.6926 & 1.9201 \\
  GDP\_PA & GDP by The Productive Approach & 0 & 0 \\
  \hline
\end{tabular}
\end{table}

\begin{table}[!t] \scriptsize
\caption{Error with malicious view $mv=0$ and manipulation factor $mf=1$} \label{table_error_1}
\centering
\begin{tabular}{ccc}
  \hline
  Method & Mean Error & Standard Deviation \\
  \hline
  CIUV & 0.9214 & 0.8455 \\
  Mean & 3.2013 & 3.8494 \\
  Median & 1.0206 & 1.1076 \\
  Voting & 1.1277 & 1.1946 \\
  3-Sources & 1.4111 & 1.5874 \\
  \hline
\end{tabular}
\end{table}

\subsection{Experimental Setting} \label{experiment_setting}
Here we describe the adopted GDP dataset, introduce the comparison methods, and define impact factors on experiment.

\it The GDP dataset. \rm We choose the GDP statistics of China in 1994 - 2014 \cite{GDP}. The GDP statistics contain 3 independent parts, statistics by the expenditure approach, income approach, and productive approach respectively. These statistics are listed in Table \ref{table_GDP_growth_rate}. We have the following relationship of statistics:
\begin{equation} \label{e_GDP}
\begin{cases}
\displaystyle GDP\_EA=FCE+GCF+NE \\
\displaystyle GDP\_IA=NPT+WC+DFA+BB \\
\displaystyle GDP\_PA=FI+SI+TI \\
\end{cases}
\end{equation}
All of these statistics can be provided as indicators for GDP growth. We take them as views of GDP growth rate (from 1995 to 2014). Since we do not actually know the most trustworthy view, we choose GDP\_PA as the grouchiest. The mean and standard deviation of the other views to GDP\_PA are also listed in Table \ref{table_GDP_growth_rate}. Note that the statistics of GDP\_FA, FCE, GCF, and NE in 2014, the statistics of GDP\_IA, NPT, WC, DFA and BB in 2004, 2008, and 2011-2014 are not given in the official site of National Bureau of Statistics \cite{GDP}. We use the GDP growth rate of their previous year to fill these blanks. Besides, the GDP dataset has a continuous value space. There exists categorical datasets under discrete value space. In this case, we can code the datasets with vectors to compute the difference between views.

\begin{table*}[!t] \scriptsize
\caption{Error comparison with varying malicious view $mv$} \label{table_mv}
\centering
\begin{tabular}{cccc|cccc}
  \hline
  \multirow{5}{*}{$mv=3$} & Method & Mean Error & Standard Deviation & \multirow{5}{*}{$mv=6$} & Method & Mean Error & Standard Deviation \\
  \hline
  & CIUV & 1.3524 & 1.2174 &  & CIUV & 2.0753 & 1.4991 \\
  & Mean & 3.3702 & 4.1237 &  & Mean & 3.7933 & 4.4065 \\
  & Median & 1.9483 & 1.3649 &  & Median & 2.7392 & 1.5693 \\
  & Voting & 1.9091 & 1.7143 &  & Voting & 2.5698 & 1.9648 \\
  & 3-Sources & 2.4202 & 2.0739 &  & 3-Sources & 2.2801 & 2.8040 \\
  \hline
  \multirow{5}{*}{$mv=9$} & Method & Mean Error & Standard Deviation & \multirow{5}{*}{$mv=12$} & Method & Mean Error & Standard Deviation \\
  \hline
  & CIUV & 2.2559 & 1.6109 &  & CIUV & 2.9077 & 1.6306 \\
  & Mean & 4.3051 & 4.6010 &  & Mean & 5.3167 & 5.4336 \\
  & Median & 3.2303 & 1.7488 &  & Median & 4.3290 & 2.8743 \\
  & Voting & 2.9949 & 1.9031 &  & Voting & 3.0689 & 1.9747 \\
  & 3-Sources & 3.3172 & 2.4766 &  & 3-Sources & 4.4670 & 1.9202 \\
  \hline
\end{tabular}
\end{table*}

\begin{figure*}[!htb] \centering
\subfigure[$mv=3$] { \label{fig_mv:a}
\centering
\includegraphics[width=1.6in]{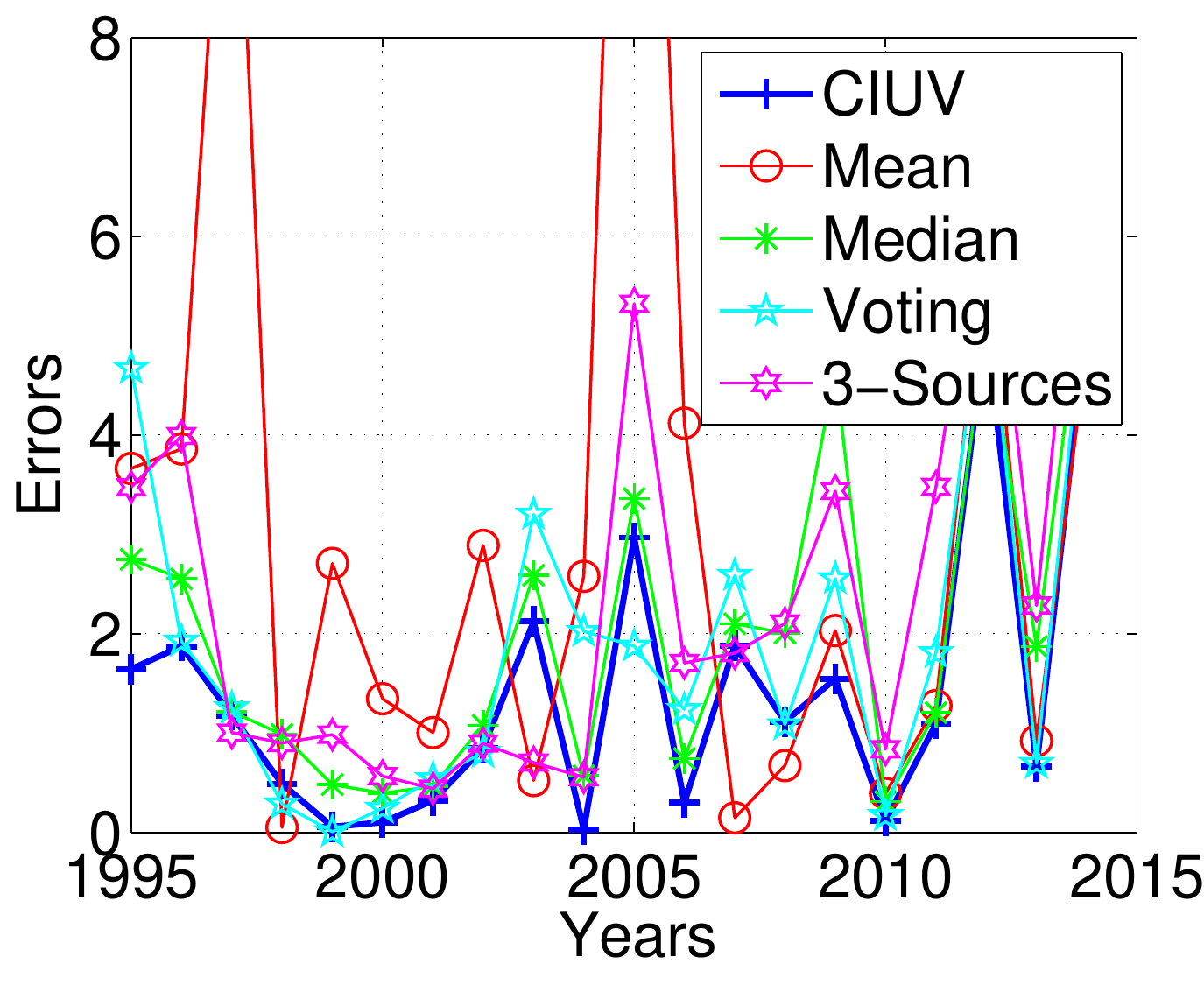}
}
\subfigure[$mv=6$] { \label{fig_mv:b}
\centering
\includegraphics[width=1.6in]{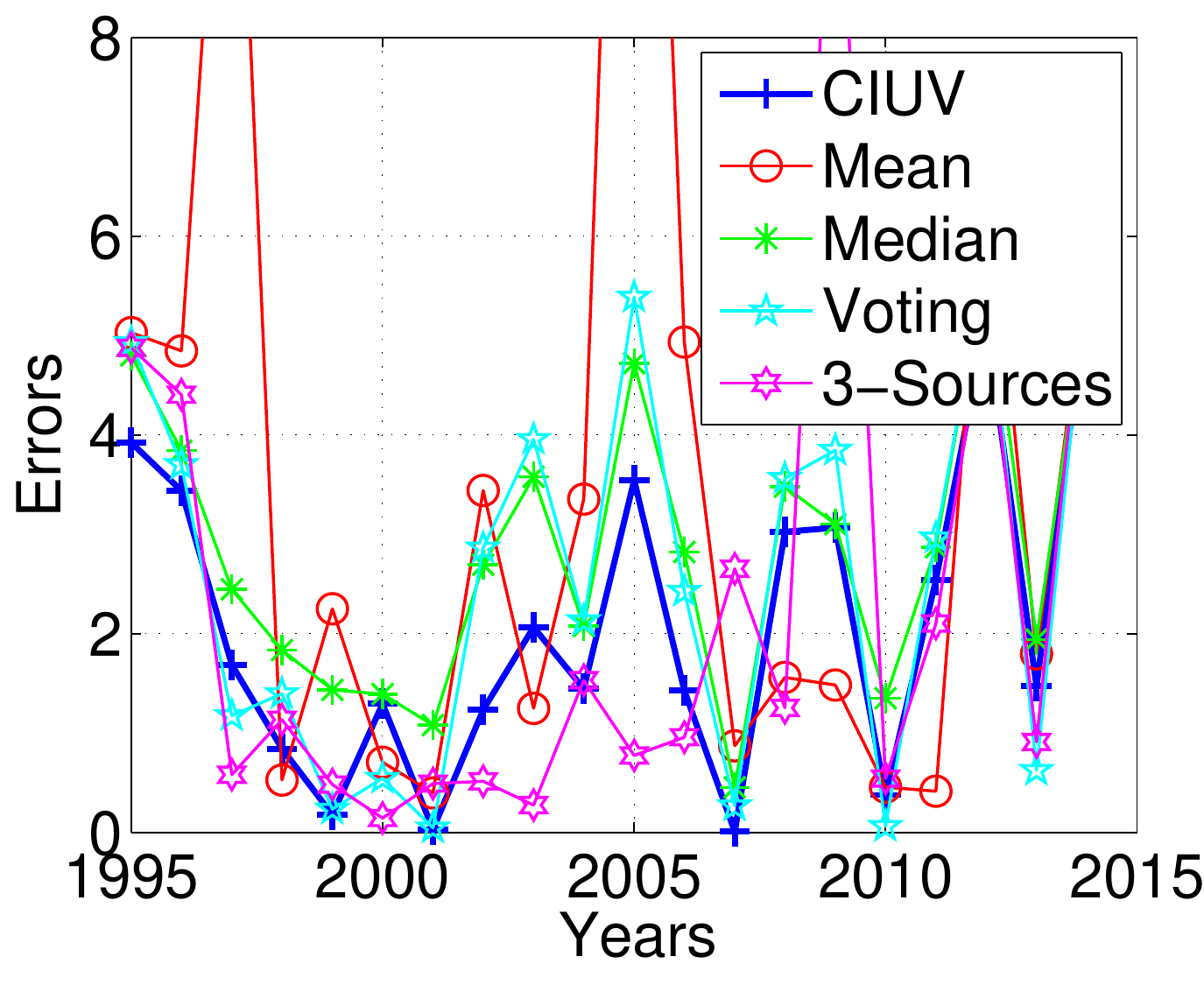}
}
\subfigure[$mv=9$] { \label{fig_mv:c}
\centering
\includegraphics[width=1.6in]{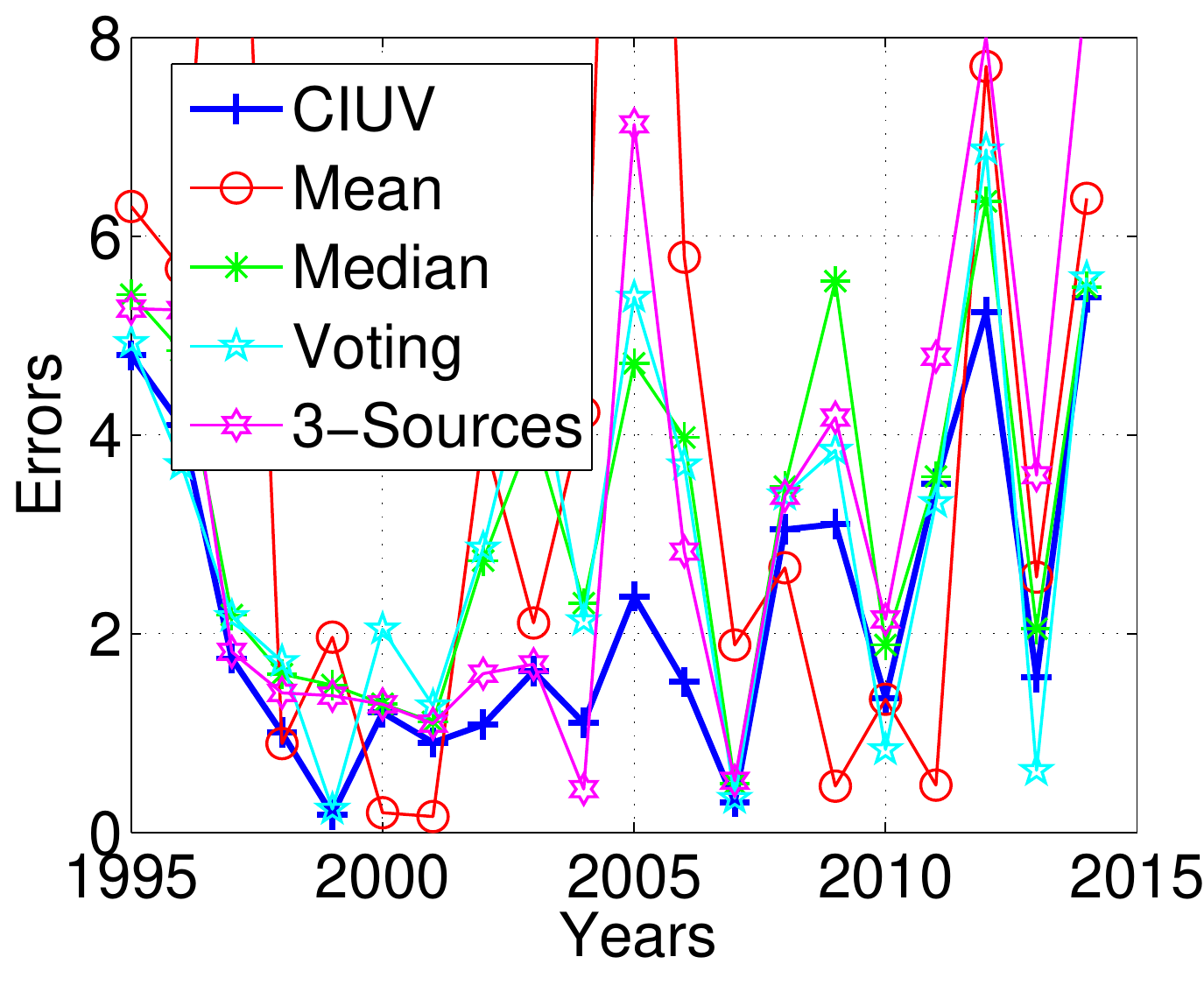}
}
\subfigure[$mv=12$] { \label{fig_mv:d}
\centering
\includegraphics[width=1.6in]{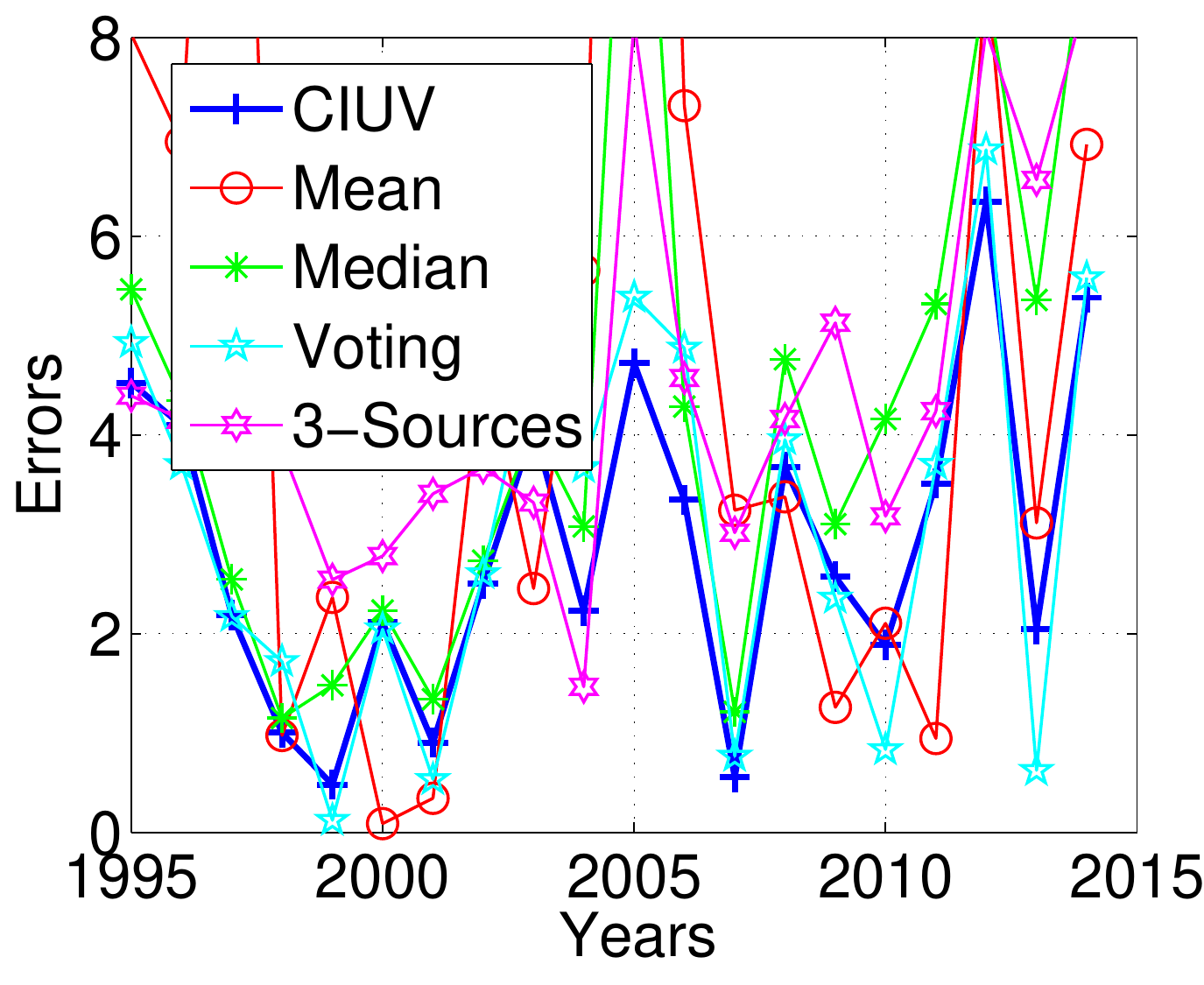}
}
\caption{Error comparison with varying malicious view $mv$}
\label{fig_mv}
\end{figure*}

\it The methods. \rm We implement four other methods besides CIUV for comparison. Two of these methods are Mean, and Median.
\begin{itemize}
  \item Mean: Use the mean value of the views $u_1,u_2,...,u_m$ to estimate $u^*$;
  \item Median: Use the median value (when $m\%2=1$) or the average of two median value (when $m\%2=0$) to estimate $u^*$.
\end{itemize}
Also, we can find too many truth-finding methods \cite{1_1}\cite{1_2}\cite{1_3}\cite{1_4}\cite{1_5} in research literatures. They are applied in different scenarios, and we cannot simply determine the best method out of them. The basic idea behind them can be mainly concluded into two categories: finding the most likely value by manipulating the data itself, and calculating the final result based on the reliability of sources. Thus we implement two methods, Voting and K-sources, in spirit related to the two categories respectively.
\begin{itemize}
  \item Voting: Let any two views $u_i$ and $u_j$ vote their distance $d(u_i,u_j)$ to each other, and choose the view nearest to all the other views as $u^*$;
  \item K-sources: Based on prior information, calculate $u^*$ by the mean value of $k$ most trustworthy views;
\end{itemize}
In our experiment, the distance $d(u_i,u_j)$ is equal to the difference of GDP growth rate from $u_i$ to $u_j$. For the K-sources method, we set $k=3$, and let the prior information be the randomly selected 10 years' statistics by default. For CIUV method, we ask 10 randomly selected problems about each year's statistics as well.

\it The experimental factors. \rm We conduct experiments on three varying factors:
\begin{itemize}
  \item Malicious View $mv$: the number of views manipulated by malicious men to prevent others from finding the truth;
  \item Manipulation Factor $mf$: the degree of multiplying factor to the original view, i.e., $u'_i=mf\cdot u_i$;
  \item Improvement Factor $if$: the degree of accuracy improvement in each iteration of CIUV process, using the following equation
      \begin{equation} \label{e_if}
        \frac{|u_i(j+1)-u_g|}{|u_i(j)-u_g|}=1-a\cdot e^{if\cdot(j+1)}
      \end{equation}
\end{itemize}

The Equation (\ref{e_if}) is raised with the intuition that the accuracy improvement by means of incentivization or punishment is fastest initially, and then slows down during iteration process. In Experiment, we set $a=0.1$. The comparison of methods with malicious view $mv=0$ and manipulation factor $mf=1$ (namely without malicious manipulation) is shown in Table \ref{table_error_1}.

\subsection{Varying Malicious View} \label{experiment_mv}
Here the experiments are conducted with varying malicious view $mv=3$, $6$, $9$ and $12$. The manipulation factor $mf$ is set to $1.2$. In experiment, the malicious views are randomly selected by 10 times for each $mv=3,6,9,12$. The corresponding results on average error from 1995 to 2014 is shown in Figure \ref{fig_mv}, and the mean and standard deviation of error comparison is shown in Table \ref{table_mv}. From the comparison results, we can see that (1) the error grows as the malicious views increases; (2) CIUV perform best among these methods (with the minimal mean and standard deviation of error for all $mv=3,6,9,12$), since CIUV considers both mean and variance of error in its design; (3) for the other methods, with informal representation, we have the average error Voting $<$ Median $<$ 3-Sources $<$ Mean. Besides, when changing the manipulation factor $mf$ under malicious view $mv=3,6,9,12$, we can get similar results.

\begin{table*}[!t] \scriptsize
\caption{Error comparison with varying manipulation factor $mf$} \label{table_mf}
\centering
\begin{tabular}{cccc|cccc}
  \hline
  \multirow{5}{*}{$mf=1.4$} & Method & Mean Error & Standard Deviation & \multirow{5}{*}{$mf=1.6$} & Method & Mean Error & Standard Deviation \\
  \hline
  & CIUV & 3.5912 & 1.7097 &  & CIUV & 3.8924 & 1.7746 \\
  & Mean & 4.9291 & 4.7829 &  & Mean & 6.2661 & 5.0928 \\
  & Median & 4.5498 & 2.1737 &  & Median & 6.3644 & 3.0368 \\
  & Voting & 4.5198 & 2.6655 &  & Voting & 6.7461 & 3.3210 \\
  & 3-Sources & 4.1925 & 2.0899 &  & 3-Sources & 5.9709 & 2.3558 \\
  \hline
\end{tabular}
\end{table*}

\begin{figure} \centering
\subfigure[$mf=1.4$] { \label{fig_mf:a}
\centering
\includegraphics[width=1.6in]{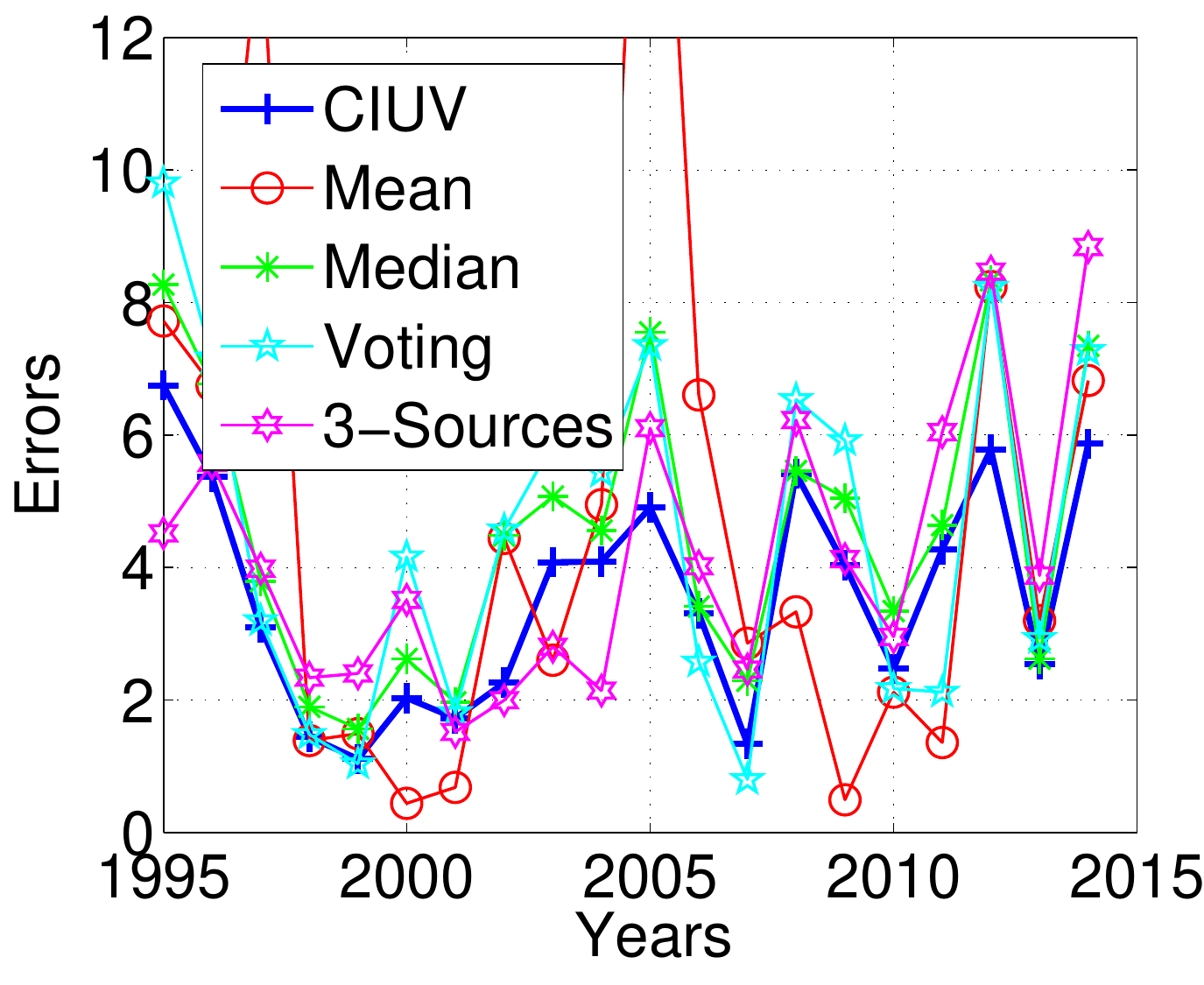}
}
\subfigure[$mf=1.6$] { \label{fig_mf:b}
\centering
\includegraphics[width=1.6in]{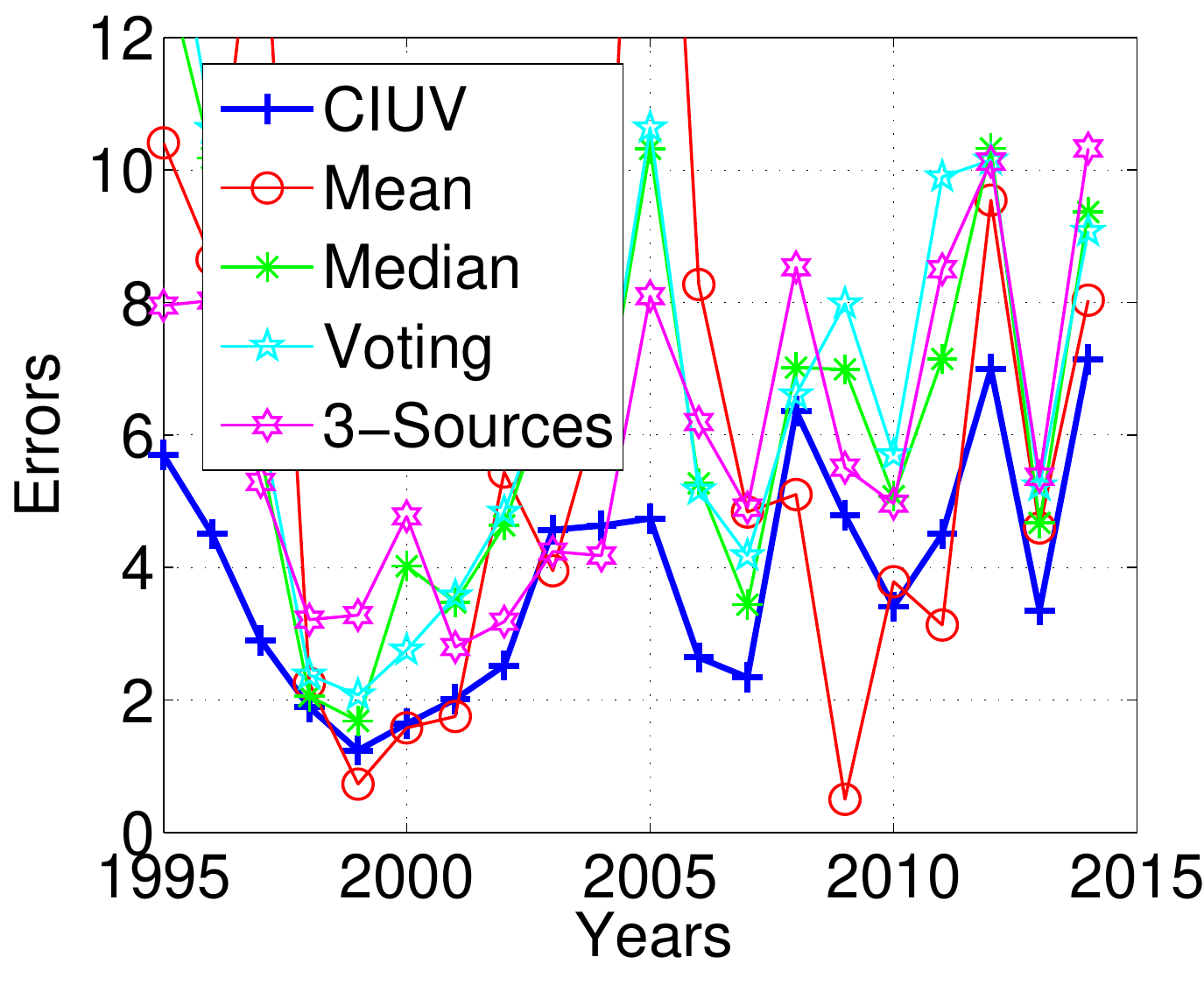}
}
\caption{Error comparison with varying manipulation factor $mf$}
\label{fig_mf}
\end{figure}

\subsection{Varying Manipulation Factor} \label{experiment_mf}
We experiment with varying manipulation factor $mf=1.4$ and $1.6$. (The case $mf=1.2$ can be seen in previous sub-section.) The malicious view $mv$, randomly selected by 10 times, is set to $6$. The corresponding results on average error from 1995 to 2014 is shown in Figure \ref{fig_mf}, and the mean and standard deviation of error comparison is shown in Table \ref{table_mf}. It can be seen that (1) the error grows drastically when manipulation factor increases compared with the change of malicious views; (2) CIUV has stable performance with the least error compared with other methods; (3) for the other methods, we have the average error 3-Sources $<$ Median $\approx$ Voting $<$ Mean.

\subsection{Varying Improvement Factor} \label{experiment_if}
\begin{figure} \centering
\subfigure[malicious view $mv=3$] { \label{fig_if:a}
\centering
\includegraphics[width=1.6in]{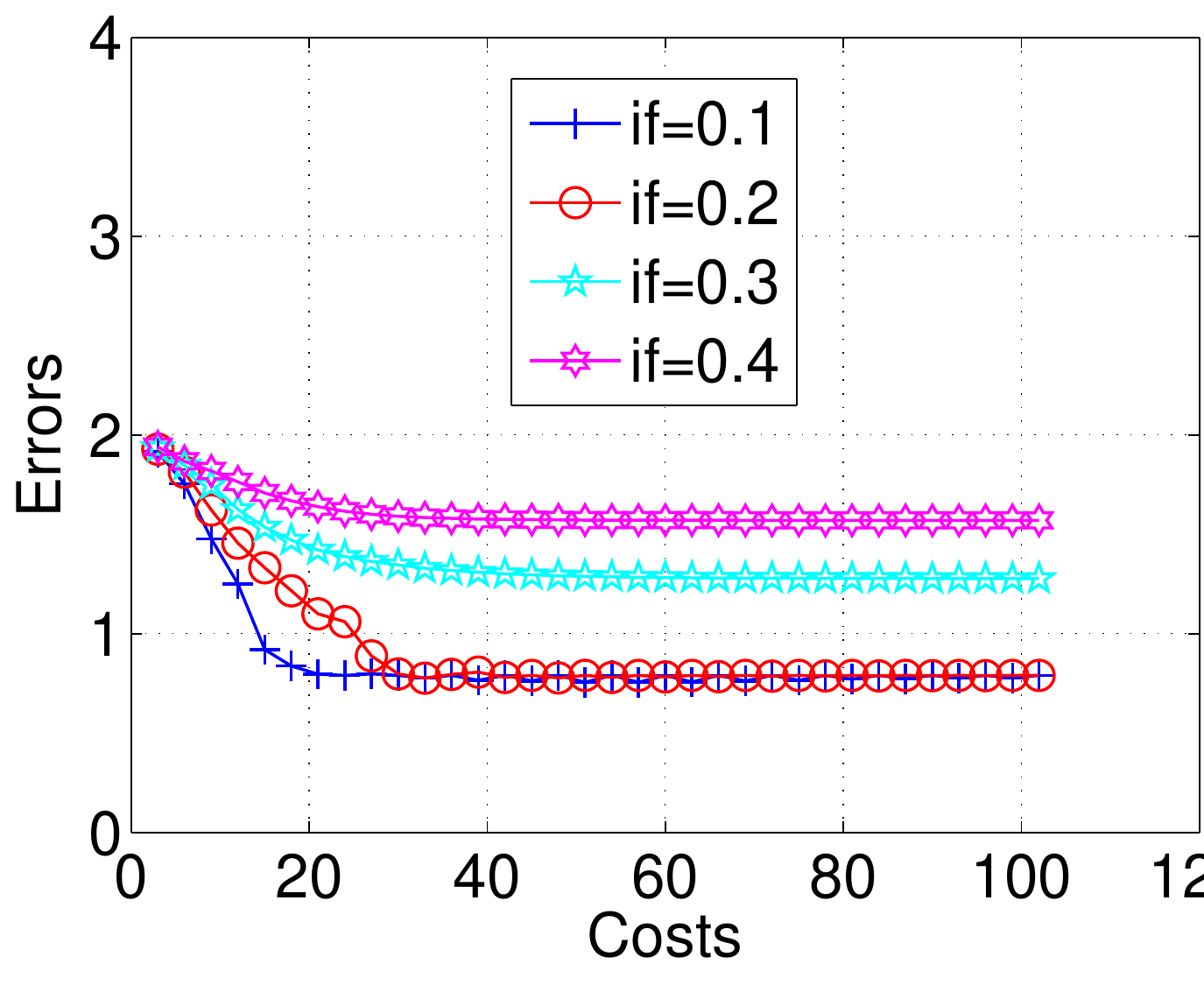}
}
\subfigure[malicious view $mv=6$] { \label{fig_if:b}
\centering
\includegraphics[width=1.6in]{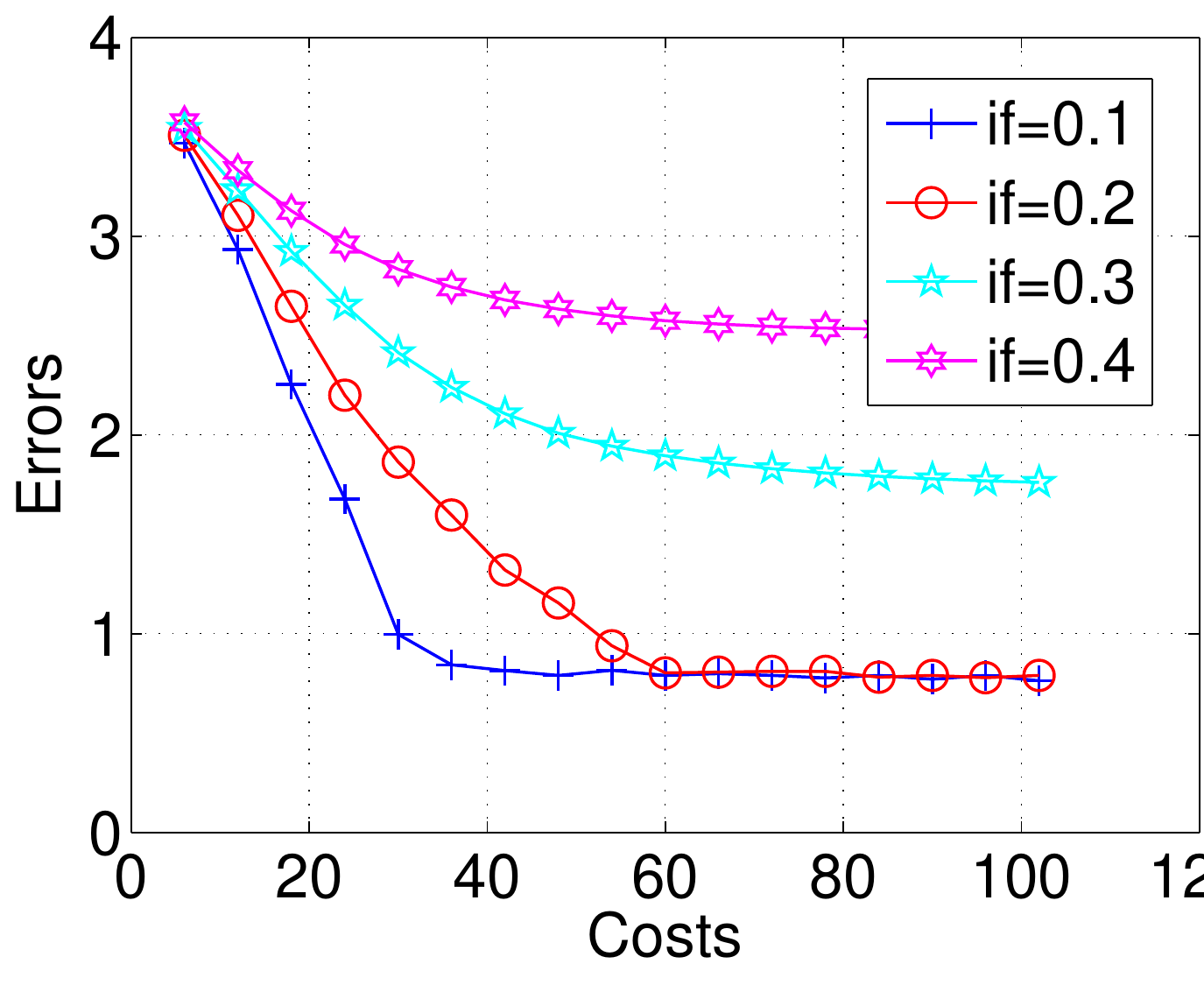}
}
\caption{Error comparison with varying improvement factor $if$}
\label{fig_if}
\end{figure}

We experiment on varying improvement factor $if=0.1$, $0,2$, $0.3$ and $0.4$. The manipulation factor $mf$ is set to $1.6$. Let the cost be the total times of effect applying on views, i.e., when malicious view $mv=3$, the times of effect is $3$ in each iteration. The results of malicious view $mv=3$ and $mv=6$ are shown in Figure \ref{fig_if:a} and Figure \ref{fig_if:b} respectively. The results show that (1) the error decreases faster when we have smaller improvement factor; (2) the costs almost double to reach a same level of stable error when changing $mv=3$ to $mv=6$. In addition, we can get similar results under other parameter settings. Here we use Figure \ref{fig_if:a} of $mv=3$ and Figure \ref{fig_if:b} of $mv=6$ as the representation of similar results.

\section{Related Work} \label{related_work}
Finding the truth on resolving conflicts from multiple sources has been studied for many years \cite{r0_1}\cite{r0_2}. An early common method \cite{ra_1}\cite{ra_2}\cite{ra_3} is to average (or to say vote) those conflicts to calculate a truth. The main idea behind it is to consider the view that appears in most times or is most similar to many other views. However, this type of method suffers from large error when there exist sources with low quality views.

To deal with the problem of low quality views, many methods were proposed to find the truth based on heuristic clues, i.e., prior knowledge on facts \cite{rb_a1}\cite{rb_a2}, source dependency \cite{rb_b1}\cite{rb_b2}, sensitivity and specificity \cite{rb_c1}. Usually, this type of method uses the clues to evaluate the reliability of sources, and calculate a truth by weighting the sources.

Learning from crowd is another related field to out work. It infers true values from the data labeled by a crowd. The methods \cite{rc_1}\cite{rc_2}\cite{rc_3}\cite{rc_4}\cite{rc_5} proposed in this research field usually focus on specific application scenarios.

\section{Conclusion and Future Work} \label{conclusion}
We propose CIUV, collaborating information to form a truth against unreliable views provided by men with malicious intension. CIUV contains 3 stages, for interactively mitigating the impact of unreliable views, and calculating the truth with the consideration of both mean and variance of error. We theoretically analyze the bound of error when implementing CIUV. In experiment, we verify the feasibility and efficiency of CIUV with varying impact factors.

In this paper, we assume that men provide their views independently. In the future, we will consider the case that men collude to provide bias views.

\end{document}